\documentclass[11pt]{article}
\usepackage{theorem}
\usepackage{latexsym}
\usepackage{amsmath}
\usepackage[latin1]{inputenc}
\usepackage[dvips]{graphicx}
\usepackage{graphicx}

\theoremheaderfont{\scshape}
\newtheorem{theorem}{Theorem}[section]
\newtheorem{corollary}{Corollary}[section]
\newtheorem{lemma}{Lemma}[section]
\newtheorem{proposition}{Proposition}[section]
\newtheorem{postulate}{Postulate}[section]
\newtheorem{statement}{Statement}[section]
{\theorembodyfont{\rmfamily}
\newtheorem{example}{Example}[section]}
{\theorembodyfont{\rmfamily}
\newtheorem{remark}{Remark}[section]}
{\theorembodyfont{\rmfamily}
\newtheorem{definition}{Definition}[section]}

\newcommand{\newc}{\newcommand}

\newc{\be}{\begin{equation}}
\newc{\ee}{\end{equation}}
\newc{\bea}{\begin{eqnarray}}
\newc{\eea}{\end{eqnarray}}
\newc{\beas}{\begin{eqnarray*}}
\newc{\eeas}{\end{eqnarray*}}

\newc{\pardt}{\partial_{t}}
\newc{\pardxi}{\partial_{i}}
\newc{\pardts}{\partial_{t^{*}}}
\newc{\pardxis}{\partial_{i^{*}}}
\newc{\pardxj}{\partial_{j}}
\newc{\pardxk}{\partial_{k}}
\newc{\pard}{\partial}

\newc{\s }{\overline}

\newc{\sect}{\section}
\newc{\subs}{\subsection}

\newc{\defi}{\definition}
\newc{\prop}{\proposition}
\newc{\rem}{\remark}
\newc{\lem}{\lemma}
\newc{\exa}{\example}
\newc{\theo}{\theorem}
\newc{\coro}{\corollary}
\newc{\post}{\postulate}
\newc{\state}{\statement}

\begin{document}
\baselineskip0.5cm
\renewcommand {\theequation}{\thesection.\arabic{equation}}
\title{Vortex length, vortex energy and fractal dimension of
superfluid turbulence at very low temperature}
\date{}

\author{\bf D.~Jou$^1$,
M.S.~Mongiov\`{\i}$^2$, 
M.~Sciacca$^{2,3}$ and C.F.~Barenghi$^3$}
\maketitle

\begin{center} {\footnotesize
$^1$ Departament de F\'{\i}sica, Universitat Aut\`{o}noma de
Barcelona,\\
08193 Bellaterra, Catalonia, Spain\\
$^2$ Dipartimento di Metodi e Modelli Matematici Universit\`a di Palermo,
\\ c/o Facolt\`{a} di Ingegneria, Viale delle Scienze, \\
90128 Palermo, Italy\\
$^3$School of Mathematics and Statistics, Newcastle University,\\
NE1 7RU Newcastle--upon--Tyne, United Kingdom}
PACS number(s):\\
 47.53.+n Fractals, fluid dynamics\\
67.25.dk Vortices in superfluid helium-4
\end{center} \footnotetext{E-mail addresses:
david.jou@uab.es (D. Jou), mongiovi@unipa.it (M. S. Mongiov\`{\i}),
msciacca@unipa.it (M. Sciacca), c.f.barenghi@ncl.ac.uk (C.F.
Barenghi). }

\begin{abstract}
By assuming a self-similar structure for Kelvin waves along vortex
loops with successive smaller scale features, we   model the
fractal dimension of  a superfluid vortex tangle  in the zero
temperature limit.   Our model  assumes that at each step the
total energy of the vortices is conserved,   but  the total length
can  change. We obtain a relation between the fractal dimension
and the exponent describing how the vortex energy per unit length
changes with the length scale. This relation does not depend on
the specific model, and shows that if smaller length scales make a
decreasing relative contribution to the energy per unit length of
vortex lines, the fractal dimension will be higher than unity.
Finally, for the sake of more concrete illustration, we relate
the fractal dimension of the tangle to the scaling exponents of
amplitude and wavelength of a cascade of Kelvin waves.
\end{abstract}

\section{Introduction}
Turbulence in helium~II, or superfluid turbulence, consists of a
tangle of quantized vortex lines \cite{Do-book-1991,BDV-book-2001}.
Until recently, in most experiments superfluid turbulence was
created in superfluid helium at rest in the presence of a heat
flux, the so-called "counterflow"  \cite{Tough,Skrbek}, an
interesting problem of non--equilibrium physics
\cite{Nemir-1998,Nemir-2006}. More recently,   superfluid
turbulence was generated by agitating the liquid helium using
grids or propellers \cite{Stalp,Tabeling,Roche}.  Particularly
interesting is the case in which the temperature $T$ is small
enough ($T < 1 \rm K$) that the normal fluid fraction of helium~II
is negligible, hence viscous dissipation and mutual friction play
no role. In this low temperature limit, superfluid turbulence
takes its purest form: a tangle of reconnecting vortex filaments
which move under the velocity field of each other.  The importance
of vortex reconnections, first recognized by Schwarz
\cite{Schwarz} and later proved by Koplik and Levine
\cite{Koplik}, cannot be underestimated \cite{scaling,Alamri}.
Vortex reconnections randomize the vortex tangle and initiate the
physical mechanisms of the decay of the tangle's kinetic energy in
the absence of viscous losses. The first mechanism is the direct
conversion of energy into sound in the form of rarefaction pulses
at reconnecting events, as predicted by the condensate nonlinear
Schroedinger equation model \cite{Adams}. The second mechanism is
a cascade of Kelvin waves of shorter and shorter wavelengths
\cite{Svistunov}--\cite{Kivotides-cascade} triggered by vortex
reconnections.  This process of generation of small scales can
proceed without significant kinetic energy losses to spatial
scales which are small enough that the kinetic energy of the
highly curved and cusped fragments of the vortices is radiated
away as sound \cite{Vinen-sound}--\cite{SvistunovPRL92(2004)} (phonon emission), that is to
say, ultimately kinetic energy becomes heat. In this regime, one
expects the vortex tangle to exhibit fractal features, if the
mentioned processes act in a self-similar way on several
orders of spatial lengths. The idea that at very low temperatures
the energy can be released by vortex reconnections, from smaller
and smaller structures, hence it increases the twisting and the
winding of the superfluid turbulence, was originally suggested a
long time ago by Feynman in his pioneering article on the
applications of quantum mechanics to liquid helium \cite{Feynman},
before we knew about fractals or the Kelvin wave cascade, and
was explored in detail by Svistunov \cite{Svistunov}.

Here we propose simple geometrical models of the fractal dimension
of superfluid turbulence, which represent reconnections and
interactions between vortex loops and the subsequent formation at
the next generation of new vortex loops and Kelvin waves on them.
The models are too simple to be dynamically realistic, but
sufficiently appealing for a qualitative understanding of some
physical features influencing the fractal dimension. We stress that
we are not attempting to develop a theory of the Kelvin wave cascade
based on actual vortex dynamics, but we shall move with simpler
considerations. Our motivations are the growing interest in
superfluid turbulence at very low temperatures
\cite{Golov-2007,Golov-2008}, and previous remarks on the fractal
nature of superfluid turbulence. In particular we remind the work of
Kivotides {\it et al.} \cite{Kivo-2002}, who numerically determined
a fractal dimension larger than unity (but at finite temperature, not in the
limit of absolute zero which we consider here), of  Nemirovskii {\it et al.}
\cite{Nemir-2002} (who considered the influence of the possible
fractal dimension of the tangle on the energy spectrum of the
turbulent velocity field) and of  Jou {\it et al.} \cite{Jou-2002} (who
proposed an heuristic form of a Vinen's generalized equation for the
dynamics of a fractal vortex tangle).

Our aim is to model  the fractal dimension of the tangle
under the condition of constant energy, but separating the scaling
behaviour of vortex length and of vortex energy during the
transfer of line length to smaller and smaller length scales. The
underlying physical idea is that the energetic contributions of
very close parts of the vortex lines may interfere with each
other, thus leading to a non additive global result for the total
energy of the loop.

First of all, we derive a general relation between the fractal
dimension of the hierarchy of self-similar vortex loops and the
behaviour of the vortex energy per unit length at different length
scales. Afterwards, in order to be more concrete and explicit, we
propose two simple models of hierarchies of self-similar loops,
whose behaviour mimics in a simplified way the features of a
cascade of Kelvin waves.

Our simple models are partially inspired to the well-known
$\beta$-model for classical intermittent turbulence
\cite{Fri-book-1995}-- \cite{Jou-Interm} and include the influence of geometrical and energetic
aspects on the fractal dimension.  We are not aware of
applications of the $\beta$-model of classical turbulence to
quantum turbulence. We think that this model can be useful to grasp some
qualitative transfer amongst different length scales. Our approach differs from that of
Svistunov \cite{Svistunov} in that it takes a less detailed, less
quantitative form, but it allows a simpler and more intuitive view
of the complicated process in question.

\section{Fractal dimension and behaviour of the vortex energy per unit length at different scales}
\setcounter{equation}{0}
Our aim is to look for an expression of the fractal dimension
$D_F$ of the vortex tangle in terms of the microscopic properties
previously mentioned, namely, vortex length distribution,
amplitude and wavelengths of Kelvin waves, and energy density per
unit length. To obtain the fractal dimension $D_F$ of the vortex
tangle,  we use the standard definition
\cite{Boder-book-1998,Manc}
\be\label{dime-frat}%
D_F=-\lim_{n\rightarrow \infty} \frac{\log (N_n/N_0)}{\log (l_n/l_0)},%
\ee
 where $N_n$ is the number of steps along a curve (or the number of
objects of a given size) and $l_n$ the length of a single step (or
the size of a given object).

We assume that the tangle can be statistically described as a
self-similar hierarchy of loops, whose forms will be discussed in
Section 3. The generation with $n = 1$ corresponds to
the level of the biggest vortices, which become more abundant
and smaller for increasing $n$. We call $N_n$ the number of vortices at the
$n$-th generation, $l_n$ the size of each loop, and $E_n'$ the energy of each loop.

Although the specific expression for the fractal dimension depends
on the details of the model, we express the fractal dimension in terms
of the energy per unit length at several scales. Note that
in our analysis $D_F$ is a property of the ensemble of
self-similar loops, not of a single loop. In fact, the
individual loops are assumed to be regular lines, and not fractal
lines. Thus, our fractal dimension characterizes the self-similarity properties of the
tangle and not of individual vortex lines.

Before proposing an explicit model of
hierarchies of vortex loops, we relate the fractal dimension
defined geometrically in (\ref{dime-frat}) with the variation of the energy
per unit length at different length scales.

We assume that $E'_n \propto l_n^{\alpha'}$, where $\alpha'$ is a
constant scaling exponent; this means that the energy per unit
length is
\be\label{1.2new}%
E'_n/ l_n \propto l_n^{\alpha'-1}.%
\ee
Therefore, if $\alpha' >1$
the contribution to the energy per unit length decreases for
lower length scales (shorter $l_n$); the opposite is true for $\alpha'
<1$. If  $\alpha' =1$ then the energy per unit length is the same
at each length scale. In principle, the exponent $\alpha'$ does not depend on
the fractal dimension (\ref{dime-frat}), but it becomes related to $D_F$ if we assume
the condition of constant total energy at the different loop
generations, as mentioned in the introduction.

According to the previous definitions, the total energy $E_n$  at
the $n$-th loop generation is given by  $E_n= N_n E'_n$. Then the
condition that the total energy is independent of $n$ can be expressed by
\be\label{1.3new}%
E_n =E_{n+1} \hskip0.4in \textrm{hence} \hskip0.4in N_n l_n^{\alpha'}=N_{n+1}l_{n+1}^{\alpha'}.
\ee
If $n$ is large enough, equation (\ref{dime-frat}) implies that $l_n
\propto N_n^{-1/D_F}$, thus equation (\ref{1.3new}) leads to
\be\label{1.4new}%
N_n^{1-(\alpha'/D_F)}= N_{n+1}^{1-(\alpha'/D_F)}. \ee

In order that this equality is true for any $n$, one must have
that  $D_F = \alpha'$. This result shows the strong connection between the energetic
features of the tangle and its geometrical structure,
independently of the detailed form of the loops in the hierarchy.
When the contribution to the energy per unit length of the smaller length scales
 is smaller than the contribution of the larger scales then $D_F>1$, whereas in the opposite case
$D_F<1$. The case $D_F<1$ seems physically unacceptable
because it would imply that vortex line fragments in objects perhaps
similar to the Cantor's dust, which would violate the condition that
the vorticity is solenoidal. This result would be supported by the numerical simulation of Kivotides
{\it et al.} \cite{Kivo-2002}.
The only acceptable situation is that the larger length scales contribute more
to the energy per unit length than the smaller length scales.
In view of the meaning of $\alpha'$, we write the fractal
dimension (\ref{dime-frat})  as
\be\label{1.5new}%
D_F-1 = \lim_{n\rightarrow \infty} \frac{\log (E'_n/l_n) }{\log
(l_n)} \ee
 which, by using  $l_n\propto N_n^{-1/D_F}$, can also be written as
\be\label{1.5new-bis}%
\frac{1-D_F}{D_F} = \lim_{n\rightarrow \infty} \frac{\log
(E'_n/l_n) }{\log (N_n)}.
\ee
In the next section, we shall introduce a model of
loop generations which relates the fractal dimension to the amplitude and the wavelength of the Kelvin
waves.

\section{Geometric and energetic assumptions}
\setcounter{equation}{0} We assume that, as the mutual interaction
of vortices induces the formation of
helical Kelvin waves along the vortex line which also undergo breaking
and reconnection processes, the tangle can be described as an
ensemble of self-similar objects. Neglecting boundaries, we assume that
these objects are closed vortex loops. Svistunov \cite{Svistunov} has
also considered the same point of view, but with different
transformation rules than those we consider here, as we shall discuss
below.

The loops can be entangled among themselves
in complex topological ways \cite{Poole_JLTP132(2003)}. Here, we
focus our attention only on geometrical properties such as vortex length, vortex number,
amplitude and wave-number of Kelvin waves, and energy per unit
length. We do not consider the topological details of the vortex
entanglement.

\begin{figure}
  \includegraphics[width=7.5cm]{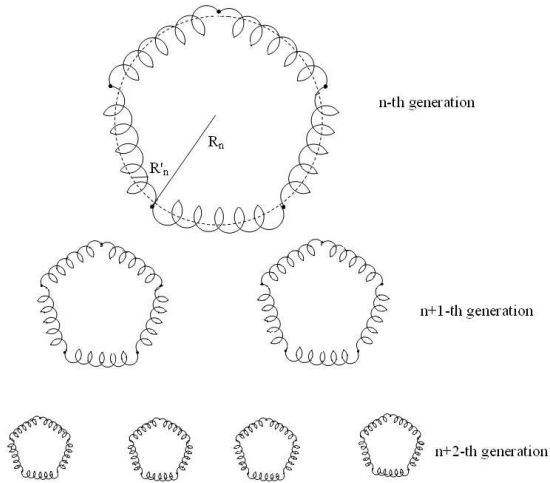}\\
  \caption{Each vortex loop at $n$-th generation with average curvature radius $R_n$
  generates $r=2$ vortex loops at $(n+1)$-th generation,
  and so on at the next generation. In the figure
  a constant number of helical turns $N'_n =N'_0$ ($\alpha=0$) is
  assumed, while the radius $R'_n$ decreases according to the rule  $R'_n =
  R'_0/\beta^{\gamma n}$. 
}\label{figura1}
\end{figure}

We envisage that the generation of vortex loops takes place according to the following rules:

\begin{enumerate}
  \item We take as reference configuration a collection of $N_0$ vortex
loops of length $L_0=2\pi R_0$, where $R_0$ is the average curvature
radius of the loop, along which  an helical structure of $N'_0$
helical turns lies, all turns being of radius $R'_0$. This structure
models in a simple way the formation of Kelvin helical waves along
vortices, where $R_0'$ is the amplitude and $h_0=L_0/N_0'$ is the wavelength of waves.
    \item The next generation is assumed to be
composed of $N_0r$ smaller vortex loops (where $r$ is a
multiplication parameter) of length $L_0/\beta$ (where $\beta$ is
another parameter). The following generation consists of $N_0 r^2$
loops of lengths $L_0/\beta^2$, and so on. Thus, the $n$-th
generation is composed of $N_n = r^nN_0$ vortex loops of lengths
$L_n = L_0/\beta^n$ (see Figure\,\ref{figura1}). Note
that the generation of smaller vortex loops at the $n$-th
generation comes from the interactions and reconnections among
vortex loops of $(n-1)$-th generation (the dots in each loop of
Figure\,\ref{figura1} denote the reconnection points). The
parameters $r$ and $\beta$ thus depend on the details of the dynamics.
    \item Besides the above set of rules which determine the number
$N_n$ of loops in the $n$-th generation and the average radius $R_n$
of the main circle of the loop, it is necessary to give a second set
of rules for $N'_n$ and $R'_n$, which are the number of helical
turns and the average radius of each turn. We interpret $R'_n$
as the amplitude of a Kelvin wave and $h_n=L_n/N_n'$ as the wavelength. We
assume that each loop has  $N'_n =N'_0 (r')^{ n}$ helical turns, and
that the radius of the helical structure at the $n$-th generation scales
as $R'_n = R'_0/\beta'^{n}$, i.e. it scales in a different
proportion than the curvature radius of the total loop. We introduce
two scaling coefficients $\alpha$ and $\gamma$, setting
$r'=r^\alpha$ and $\beta'=\beta^{\gamma}$ (in Figure \ref{figura1},
for the sake of simplicity, it is assumed that the number of helical
turns in each vortex loop is constant, that is $\alpha=0$).
\item  In the breaking and reconnection processes the total energy
remains unchanged because at the considered length scales there is
no friction, hence no energy dissipation. Thus, we impose that $E_{n+1} = E_n$ for all values of
$n$. This assumption has a limiting length scale because at some very small length scale
sound radiation becomes relevant and the fractalization process stops.
\end{enumerate}

We stress that the first three rules are not supposed to model actual
dynamical processes; they are only meant to explore the possible
consequences of two physical processes which, in this context of
superfluid turbulence at very low temperatures, are still poorly
understood: a direct cascade, leading from bigger to smaller
loops, and an inverse cascade, leading from smaller to bigger
loops.
Concerning the fourth step, another plausible choice could be the invariance of the total
length, instead on the invariance of the energy;  this provides another way to
examine the fractal dimension which leads to $D_F=1$. This result follows
from the general arguments of the Section 2. The assumption that
the total length of the vortex line is the same at each step $n$ means
\be\label{leng_invariante}%
N_n l_n=N_{n+1} l_{n+1}.%
\ee
Equation (\ref{dime-frat}) implies that $l_n
\propto N_n^{-1/D_F}$ for $n$ high enough, so equation (\ref{leng_invariante}) becomes
\be\label{leng_invariante1}%
{N_n}^{(1-1/D_F)}=N_{n+1}^{(1-1/D_F)}.%
\ee
Since relation (\ref{leng_invariante1}) is true for every $n$ high enough, then $D_F=1$.
In our opinion it is an interesting result because it shows that
the fractal dimension is 1 if the whole vortex line length is kept constant, that is
if vortices themselves do not contribute to lengthen o shorten of the total vortex length.
But, as pointed out along this paper, in the low temperature limit it seems more
plausible the invariance of the total energy with respect to the invariance of the total length.

Since the total number of vortex loops at the $n$-th generation is
$N_n$, the total energy stored on the vortex loops of the $n$-th
generation will be
\be\label{ene-loop-En2}%
E_n=N_n E'_n%
\ee
where $E'_n$ is the energy of the loop at the $n$-th generation.
The length of a single vortex in the $n$-th generation is
\be\label{lenght-n-elica}%
l_n=2\pi N'_n R'_n\sqrt{1+\left(\frac{R_n}{R'_nN'_n} \right)^2 }
=2 \pi N'_0 R'_0 \frac{r^{n \alpha  }}{\beta^{n\gamma}} \sqrt{
1+\left(\frac{R_0}{N'_0 R'}\right)^2
\left(\frac{\beta^{\gamma-1}}{r^\alpha} \right)^{2n}}.%
\ee
Finally, we need to specify the energy of each loop. Of course, the energy of a loop
depends on the helical structure that wraps the
unperturbed loop of average radius $R_n$, that is, on the number
of helical turns $N'_n$ or, equivalently, on the pass of helices.
Unfortunately we are unable to calculate this energy analytically. The recent work of
Maggioni {\it et al.} \cite{Maggioni_2009} shows that, even for the simpler case
of a non fractal, single vortex filament, the numerical calculation of this energy is
difficult, as it converges very slowly.
For this reason, we propose
three possible approximate scenarios for the energy $E'_n$
of a loop at the $n$-th generation.

\begin{enumerate}
  \item  The first scenario assumes that $E'_n$ is proportional
to the length of the circular axis of the helical loop:
\be\label{ene-loop}%
E'_n=\frac{\rho_s \kappa^2}{4 \pi}L_n\left[\log
\left(\frac{8R_n}{a_0}\right)-1.615\right],%
\ee
where $\rho_s$ is the mass density of the superfluid
component, $\kappa$ is the quantum of vorticity ($\kappa = h/m$,
with $h$ Planck's constant and $m$ the mass of the helium atom), and
$a_0$ is the radius of the core of the vortices (of the order of the
atomic radius) \cite{Do-book-1991}.

\item The second scenario is to assume
that for large-amplitude helical turns close to each
other ($R'_n \gg h_n$) the helical loop may be considered as
a solenoid, and that its internal energy $E'_n$  is of the order
of its volume, $2\pi R_n\pi {R'_n}^2$, times the density of the
kinetic energy of the superfluid induced by the polarized coil.
Using for the induced velocity an expression analogous to that for
the magnetic field in a solenoid we can write for the induced
velocity $v_{sl}\approx \kappa/h_n$ and obtain
\be\label{3.1new}%
E_n' = \pi^2 \rho_s \kappa^2 R_n \left( {R_n' \over h_n}\right)^2. %
\ee

\item The third scenario is a more flexible prescription, incorporating
also the features of the helical structure.
The total length of the deformed circle, for $n$
sufficiently high, is
\be\label{length-loop}%
l_n=2\pi N'_n R'_n\sqrt{1+\left(\frac{h_n}{2\pi R_n'} \right)^2},
\ee
with $h_n=L_n/N'_n$ being the pass of the helices at the $n$-th
generation, which can be interpreted as the wavelength of the Kelvin
wave, whose amplitude is $R'_n$.
We assume that the whole
helical length contributes to the energy, but multiplied by a
dimensionless factor depending on the length scale, as represented,
for instance, by $R'_n$, which modulates the relative influence of
the helical turns on the energy of the loop. Thus, instead of
(\ref{ene-loop}),  we take
\be\label{ene-loop-alt}%
E'_n=\frac{\rho_s \kappa^2}{2} N'_n R'_n
\left(\frac{R'_n}{R_0}\right)^\chi \sqrt{1+\left(\frac{h_n}{2\pi
R'_n} \right)^2 } \left[\log
\left(\frac{8R'_n}{a_0}\right)-\delta\right],%
\ee where $\delta$ is a constant of the order of $1.6$.
 Here the ratio $(R'_n/R_0)^\chi$ with $\chi>0$  ensures that
the smaller the length scale is (i.e. the smaller $R'_n$ is), the
smaller is the contribution to the energy;  if $\chi<0$, smaller
lengths scales have larger contributions to the energy, and $\chi=0$
indicates that the energy is proportional to the loop length. Of
course, more complicated models could be assumed instead of
$(R'_n/R_0)^\chi$. The actual value of $\chi$ should be obtained
from a first principles calculation of the energy of helical loops
of different radii and with different separations between successive
helical turns, but, as we said before, this is very complicated and goes beyond the
simple, limited task which we set at this early stage of
investigation.
\end{enumerate}
Since  $N_n/N_0 = r^n$ and $l_n/l_0=l_n/2\pi R_0$, when $l_0$ is assumed to be $2\pi R_0$,
 the fractal dimension (\ref{dime-frat}) becomes
 \be\label{dime-frat2}%
D_F=-\lim_{n\rightarrow \infty} \frac{n\log r} {\log
\left(\left(N'_0 R'/R_0\right)( r^{\alpha n}/\beta^{\gamma n})
\left[1+\left[R_0/(R' N'_0)\right]^2
\left({\beta^{\gamma-1}}/{r^\alpha} \right)^{2n}\right]^{1/2}
\right)}. %
\ee
Note that here the limit is taken keeping in mind  that, when $n$
becomes large enough, the radius of the loop $R_n$ cannot be smaller,
or of the same order of  the vortex core radius $a_0$.
We  distinguish
essentially two cases od physical interest: the long wavelength limit ($h_n\gg R_n$) and
the large amplitude limit
($h_n\ll R_n$), corresponding respectively  to Kelvin waves
whose wavelengths are larger or smaller than  their amplitudes.

\subsection*{Long wavelength limit ($h_n\gg R_n'$ or $\beta^{\gamma-1}>r^\alpha$)}
In the limit of Kelvin waves with small amplitude and large wavelength,
the physically most plausible scenarios are the first (\ref{ene-loop}) and the third (\ref{ene-loop-alt}).

In the first scenario (\ref{ene-loop}),  under the condition $E_{n+1} = E_n$, leads to
\be\label{condi_r_beta}%
\left[\left(n \ln
\beta + a \right) \right]
 ={r\over \beta}
 \left[\left((n+1) \ln \beta + a \right) \right],%
\ee
which is true for any value of $n$ only if $r=\beta$ needs.
It follows that the condition $h_n\gg R_n$, or $\beta^{\gamma-1}>r^\alpha$
can be read in terms of $\gamma$ and $\alpha$ as
$\gamma-\alpha>1$, and the value of the fractal dimension can be obtained from (\ref{dime-frat2}):
\be\label{dim_caso1_ene1}%
D_F=1 \qquad \textrm{if} \qquad \gamma-\alpha>1.%
\ee
It is reasonable that $D_F = 1$ because in this scenario the
interference between neighboring helical turns  tends to vanish.

In the third scenario, multiplying (\ref{ene-loop-alt}) times $N_n$ and requiring
that $E_{n+1}=E_n$, i.e., $N_{n+1}E'_{n+1} = N_nE'_n$, we
are lead to
\[
\frac{r^{(\alpha+1)(n+1)}}{\beta^{\gamma(\chi+1)(n+1)}} \sqrt{
1+\left(\frac{R_0}{N'_0
R'_0}\right)^2\left(\frac{\beta^{\gamma-1}}{r^\alpha}
\right)^{2(n+1)}} \left[-(n+1)\gamma \log \beta + a \right]= \]
\be\label{cond-new}
=\frac{r^{(\alpha+1)n}}{\beta^{\gamma(\chi+1)n}}
 \sqrt{
1+\left(\frac{R_0}{N'_0
R'_0}\right)^2\left(\frac{\beta^{\gamma-1}}{r^\alpha}
\right)^{2n}}
 \left[-n\gamma\log \beta + a \right].%
 \ee
The second term under the square root in
(\ref{cond-new})  is dominant and
the relation between $r$ and $\beta$ becomes
\be\label{beta-alpha-infty}%
r= \beta^{ 1+\gamma\chi }.
\ee Substituting this relation into $\beta^{\gamma-1}/r^{\alpha}>1$
one gets $(\gamma-\alpha-\alpha\gamma\chi-1)/((1+\gamma \chi))>0$.

Using (\ref{lenght-n-elica}) and $N_n/N_0= r^n$ we have
\be\label{dime-frat-alt-infty}%
D_F= {1+\chi\gamma} \geq 1 \qquad \textrm{if} \qquad \chi\geq 0. %
\ee    Note that a negative  value of $\chi$
implies  $D_F<1$, and that the value of $\chi$ cannot be less
than $-1/\gamma>-1$. The result that $\gamma>1$ comes from the relation $(\gamma-\alpha-\alpha\gamma\chi-1)/((1+\gamma \chi))>0$.
The conclusion that $D_F<1$ is not physically reasonable: it  would
imply that the vortex tangle becomes similar to a Cantor's dust and would violate the solenoid condition
(a vortex is a closed loop or terminates on boundaries, but here we have no boundaries).

For the sake of completeness, we also consider scenario two (\ref{3.1new})
although this expression of the energy
seem to be physically  inadmissible in the long wavelength limit. The constraint $E_{n+1} = E_n$ leads
to $r^{1+2\alpha}=\beta^{2\gamma-1}$ and  $\gamma-\alpha-1>0$. Substitution into
(\ref{dime-frat2}) yields
\be%
D_F=\frac{2\gamma-1}{2\alpha+1}> 1,%
\ee
(as $D_F-1$ is positive when  $\gamma>\alpha+1$).

\subsection*{Large amplitude limit ($h_n\ll R'_n$ or $\beta^{\gamma-1}<r^\alpha$)}
Suppose that Kelvin waves have amplitude larger
than the wavelength. This means that the helical turns which wrap loops at the $n$-th generation
must have radius $R'_n$ larger than their pass $h_n$. The most plausible assumptions
for the energy seems scenario  (\ref{3.1new}) and
(\ref{ene-loop-alt}).

Using  (\ref{3.1new}) and $E_{n+1} = E_n$ we obtain
\be\label{condi_r_beta2}%
\left[\pi \left( {r'_0N'_0\over L_0} \right)^2 \left(
{r^\alpha\over \beta^{\gamma-1}} \right)^{2n} \right]
 ={r\over \beta}
 \left[ \pi \left( {r'_0N'_0\over L_0} \right)^2
\left( {r^\alpha\over \beta^{\gamma-1}} \right)^{2n+2}\right],%
\ee
which is  valid at any generation $n$ if $r^{1+2\alpha}=\beta^{2\gamma-1}$.
Substitution into (\ref{dime-frat2})  gives
\be%
D_F=\frac{2\gamma-1}{\alpha+\gamma}<1,%
\ee
provided that $\gamma-\alpha<1$, which, as we have already said, seems physically unplausible.

Using (\ref{ene-loop-alt}) we obtain again (\ref{cond-new}), but here with the condition
\be\label{beta-alpha}%
\beta^{\gamma(1+\chi)}=r^{\alpha+1}.%
\ee
Substitution into the inequality $\beta^{\gamma-1}/r^{\alpha}<1$ yields
 $(\gamma-\alpha-\alpha\gamma\chi-1)/(\gamma(1+\chi))<0$.

Then, substituting (\ref{beta-alpha}) into (\ref{dime-frat2}), we obtain
\be\label{dime-frat-alt}%
D_F=\frac{1+\chi}{1-\alpha \chi},%
\ee
which implies that
\be
\frac{D_F-1}{D_F}=\frac{\chi(1+\alpha)}{(1+\chi)}
\ee
making apparent that
\be\label{caso1}%
D_F\geq 1  \qquad \textrm{if} \qquad \chi\geq 0,%
\ee
(otherwise $D_F<1$). The last conclusion requires
$-1<\chi<0$, because $0<1/D_F<(1+\alpha)/(\gamma(1+\chi))$.

Again, although scenario one seems
to be unphysical, we remark for the sake of completeness that the
condition $r=\beta$, obtained below equation (\ref{condi_r_beta}), leads to
\be\label{dime-frat5}%
D_F=\frac{1}{\gamma-\alpha}>1
\ee
provided that $\gamma-\alpha < 1$.

\section{Concluding remarks}\setcounter{equation}{0}

Finally, the following two comments are worthwhile. The first is
that, at sufficiently low temperatures, the energy-conserving
process that breaks or lengthen vortices does not continue
indefinitely, but terminates at sufficiently small scales,
where a significant amount of energy is dissipated as sound. The
dependence of this energy radiation  upon the length scale has
been studied by Vinen \cite{Vinen-sound}, and Kozik and B. Svistunov
 \cite{SvistunovPRL94(2005)}--\cite{SvistunovPRL92(2004)}. According to the
Vinen's analysis, sound radiation becomes relevant at length scales of the
order of $l_{min}\simeq(\kappa^3/\epsilon)^{1/4}$, where $\kappa$
is the quantum of  circulation and $\epsilon$ is the energy
communicated to the system per unit volume and time, which is
proportional to $L^2$. Thus, $l_{min}$  is proportional to
$L^{-1/2}$. Thus sound emission limits the Kelvin wave cascade
process considered here.  In classical turbulence, viscous
dissipation plays a similar role, and determines the smallest
scale $l_{diss}\sim [\nu^3/\epsilon]^{1/4}$ for which the
celebrated Kolmogorov scaling is valid
\cite{Fri-book-1995}-\cite{Boder-book-1998}.

Second, the correlation between Kelvin waves has not been
considered in this simple model; in \cite{Nemir-2002}, however, it
has been argued it could play a significant role in the fractal
properties; it would be interesting to explore their contribution
in a more detailed model. Our model is too simple to do so; it
must be stressed that reconnections play a decisive role in the
breaking bigger loops into smaller loops; without reconnections,
energy and momentum conservation would forbid this cascade
\cite{Svistunov}.

In summary, we have proposed some toy models which allow
to interpret the fractal dimension of a vortex tangle in energetic
terms. Their energy is not proportional to the vortex length
--- because of the mutual interference of very close parts of the
vortex line --- and energy, rather than the length, is conserved
in the breaking and recombination of vortices. For example, very recent work by Maggioni {\it et al.}
\cite{Maggioni_2009} has demonstrated that for complex vortex structures such as vortex
coils and vortex knots, the energy per unit length is not constant; a
similar effect may occur on vortex filaments in superfluid turbulence. We have determined
a relation between the fractal dimension and the influence of the
smaller length scales on the total energy. If this influence is
smaller than that of the bigger length scales, the fractal
dimension is higher than 1. This result can be understood in an
intuitive way, because energy restrictions
 do not restrict the presence of many small and complicated
vortex loops,  which tend to fill a proportion of space higher
than a simple geometrical line. In contrast, if smaller scales
contribute considerably to the energy, energy restrictions limit
the formation  of these scales and vortex loops will be relatively
large and simple. Our results show that when small length scales
contribute relatively less to the energy than the long scales, the
fractal dimension $D_F$ is larger than 1. The opposite is not
true; one could have a fractal dimension higher than 1, but with
an essentially linear relation between  energy and length. The
logarithmic dependence of the fractal dimension on the behavior of
the energy per unit length at different scales may allow to obtain
reasonably physical result for $D_F$ without knowing in full
detail the exact form of the energy contribution of loops at
different scales.

Another interesting result is pointed out in Section 3, and it regards what
happens if the whole vortex line length is kept invariant at each generation $n$, instead
of the total energy. The result is that the fractal dimension
has to be 1, that is the assumption that each vortex does not contribute to the
lengthening or shortening of the other vortices means that vortices are not
fractals.

\section*{Acknowledgments}
The authors acknowledge the support of the University of Palermo
(Progetto CoRI 2007, Azione D, cap. B.U. 9.3.0002.0001.0001). DJ
and MSM acknowledge the collaboration agreement between
Universit\`{a} di Palermo and Universit\`{a}t Aut\`{o}noma de
Barcelona. DJ acknowledges the financial support from the
Direcci\'{o}n General de Investigaci\'{o}n of the Spanish Ministry
of Education under grant Fis2006-12296-c02-01 and of the
Direcci\'{o} General de Recerca of the Generalitat of Catalonia,
under grant 2009 SGR-00164. MSM and MS acknowledge the financial
support of the Universit\`{a} di Palermo under grant  2006
ORPA0642ZR. MS acknowledges the financial support of INDAM.
 CFB acknowledges the support of the
EPSRC.

\end{document}